# Fission rate of excited nuclei at variable friction in the energy diffusion regime

**M V Chushnyakova[1], I I Gontchar[2]**

[1]Physics Department, Omsk State Technical University, Omsk 644050, Russia
[2]Physics and Chemistry Department, Omsk State Transport University, Omsk 644046, Russia

E-mail: maria.chushnyakova@gmail.com

**Abstract.** Presently, it is well established that fission of excited nuclei is a dynamical process being a subject of fluctuations and dissipation. In the literature, there are indications that, at the compact nucleus shapes, the strength of nuclear friction is significantly reduced in comparison with the prediction of the one-body approach. Thus, one can expect that at small deformations the nuclear fission process occurs in the so-called "energy diffusion regime". The purpose of our present work is to compare an approximate analytical formula for the fission rate in this regime with the quasistationary numerical rate which is exact within the statistical errors. Our calculations demonstrate relatively good agreement between these two rates provided the friction parameter is deformation independent. If one accounts for its deformation dependence, the agreement becomes significantly poorer.

## 1. Introduction
The problem of nuclear friction in the fission process was first addressed by Kramers [1] in 1940 and still remains open [2–5].

On one hand, there seems to be a consensus that its physical origin is the one-body dissipation [5–7]. However, there are some indications that the strength of friction is significantly reduced when the nuclear shape is compact [8–13].

Thus, one can expect that the nuclear motion at small deformation happens in the regime which is called in the literature "the energy diffusion regime" [1,14–16]. An approximate analytical formula for the fission rate in this regime, obtained in [1] and modified in [17], was not carefully compared with the exact quasistationary numerical rate so far. The purpose of our contribution is to perform such a comparison.

## 2. Model
For this aim, we model the fission process using the Langevin equations. Since in [1,17] only one-dimensional (1D) formula for the fission rate was derived, we perform our modeling for 1D too. The motion of the fissioning nucleus is described by a dimensionless coordinate $q$ which corresponds to nuclear elongation and by the conjugate momentum $p$. In the discrete form the equations used for the present modeling read:

$$p^{(n+1)} = p^{(n)}\big(1 - \eta^{(n)} m^{-1}\tau\big) + K\tau + g^{(n)} b^{(n)} \sqrt{\tau}, \tag{1}$$

$$q^{(n+1)} = q^{(n)} + \big(p^{(n)} + p^{(n+1)}\big)\tau/(2m). \tag{2}$$

The superscripts represent two consequent time moments separated by the time step of numerical modeling $\tau$. The random numbers $b$ entering the random force have a Gaussian distribution with zero average and variance equal to 2. In equation (1), $\eta(q)$ is the coordinate dependent friction parameter; $K = -dU/dq$ is the driving force; $g(q) = \sqrt{\theta\, \eta(q)}$ is the amplitude of the random force; $\theta$ stands for the temperature of the excited nucleus; $m$ is the inertia parameter.

Recently, we proposed to model the process in this regime by means of a Langevin equation for the action as the stochastic variable [16]. This approach is more efficient with respect to the time of computer modeling, however, the approach based on equations (1), (2) is more accurate though more computer time-consuming.

The potential $U(q)$ is represented by two parabolas of the same stiffness $C = 66.8$ MeV smoothly jointed at $q = q_m$:

$$U(q) = \begin{cases} C(q - q_c)^2/2 & \text{at } q < q_m; \\ U_b - C(q - q_b)^2/2 & \text{at } q > q_m. \end{cases} \quad (3)$$

Here "$c$" refers to the ground state and "$b$" corresponds to the top of the barrier ($q_b = 1.6$). At the initial moment of time, the particle is at rest near the minimum of the potential well with the coordinate $q_c = 1.0$.

The modeling results in a sequence of $N_{tot}$ trajectories terminated not later than at the time moment $t_D$. Some of the trajectories reach the absorptive border $q_a = 2.0$ before $t_D$. The fission rate is calculated in this algorithm as follows

$$R_{at}(t) = \frac{1}{N_{tot} - N_{at}} \frac{\Delta N_{at}}{\Delta t}. \quad (4)$$

Here $N_{at}$ is the number of fictious Brownian particles (fissioning nuclei) which have reached $q_a$ by the time moment $t$, $\Delta N_{at}$ is the number of particles arriving at $q_a$ during the time interval $\Delta t$. Two examples of $R_{at}(t)$-dependence are shown in figure 1 for the case of the deformation independent friction. One sees that after some relaxation time, the rate becomes approximately time-independent. To find this quasistationary rate $R_D$ we choose several bins beginning from the end of $R_{at}$-array and make averaging over these bins. This procedure and its errors are discussed in detail in [16,18].

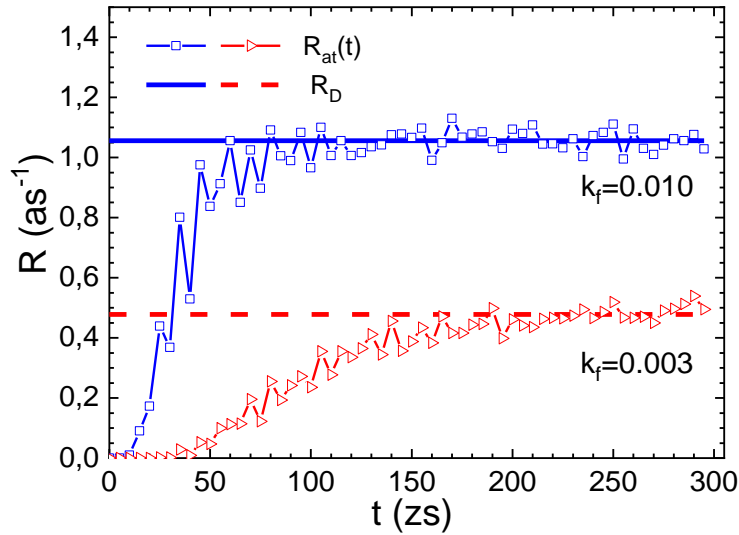

**Figure 1.** Two time-dependent fission rates calculated for the deformation independent friction parameter (wriggling lines with symbols) and their quasistationary values (thick horizontal lines) for two values of the friction strength parameter $k_f$ (see equation (5)).

For the case when the friction parameter does not depend upon the deformation, it is evaluated as

$$\eta = k_f \eta_0 \tag{5}$$

with $\eta_0 = 460$ MeV · zs. In the case of the deformation dependent friction parameter, at $q > q_c$ we apply the following formula

$$\eta(q) = k_f \eta_0 \left\{1 + S_\eta \left[1 - \exp\left(-\frac{2(q-q_c)^2}{(q_b - q_c)}\right)\right]\right\} \tag{6}$$

whereas at $q < q_c$ we use equation (5). The dimensionless parameter $k_f$ serves for varying the absolute value of friction; parameter $S_\eta$ allows varying the strength of $\eta(q)$-dependence. This behavior of $\eta$ as a function of $q$ is illustrated in figure 2.

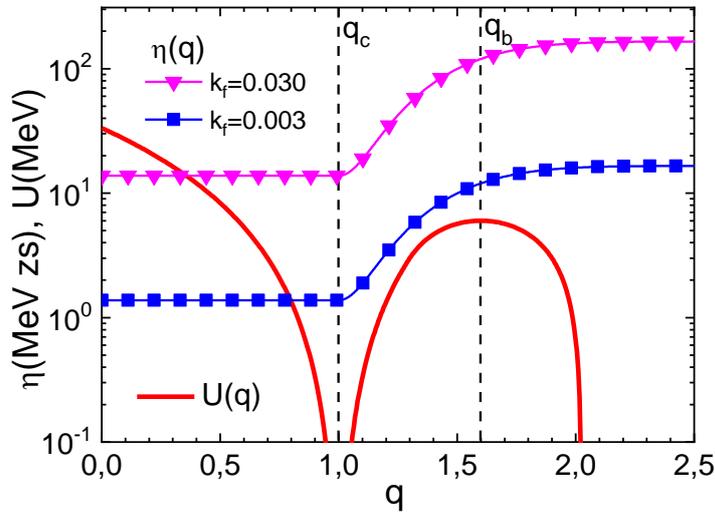

**Figure 2.** The coordinate dependence of the friction parameter according to equation (6) used in the present work (curves with symbols, $S_\eta = 11$) and the potential energy (curve without symbols). The vertical lines indicate the positions of the ground state and the barrier.

Let us now discuss the approximate formula for the quasistationary fission rate when there is almost no dissipation in one bounce. The formula was obtained in [1]

$$R_{KL} = \gamma \frac{\omega}{2\pi} \exp\left(-\frac{U_b}{\theta}\right). \tag{7}$$

Here $\omega = \sqrt{C/m}$ and

$$\gamma = \frac{I_b \eta}{\theta m}; \tag{8}$$

$I_b$ denotes the classical non-dissipative action at the collective energy equal to $U_b$. In the case of the applied potential profile,

$$I_b = 2\pi \cdot 1.07 \cdot \frac{U_b}{\omega}. \tag{9}$$

Equation (7) is supposed to be valid provided $\gamma < 1$. All calculations in this work are performed with $U_b = 6$ MeV, $\theta = 1.5$ MeV, $m = 122$ MeV zs$^2$.

The rate $R_{KL}$ corresponds to the so-called energy diffusion regime when $\eta$ primarily provides fluctuations. In [17], a modification of equation (7) was proposed allowing a smooth transition from the energy diffusion regime to the phase space diffusion regime. Denoting this modified rate as $R_{KLB}$ we write it as

$$R_{KLB} = \frac{\delta - 1}{\delta + 1} R_{KL} \quad (10)$$

with

$$\delta = \left(1 + \frac{4\alpha}{\gamma}\right)^{1/2}; \quad (11)$$

$\alpha$ is a dimensionless adjustable parameter of the order of unity.

### 3. Results

To compare the numerical rate $R_D$ with the approximate $R_{KLB}$, we first have to adjust the value of $\alpha$ in equation (11). In figure 3(a), we show the rates obtained with the coordinate-independent friction parameter. The circles indicate results of dynamical modeling. The statistical errors estimated as the double root-mean-square deviation in the present modeling do not exceed 4%. The values of $R_{KLB}$ calculated using equation (10) with $\alpha = 1.0$ are shown in figure 3(a) by the solid curve without symbols. One sees rather good agreement between the approximate rates $R_{KLB}$ and the numerical rate $R_D$ (we believe the latter is exact within the statistical errors).

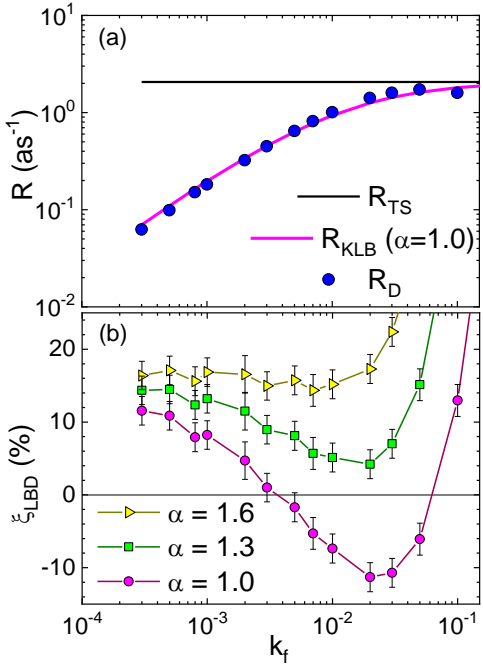
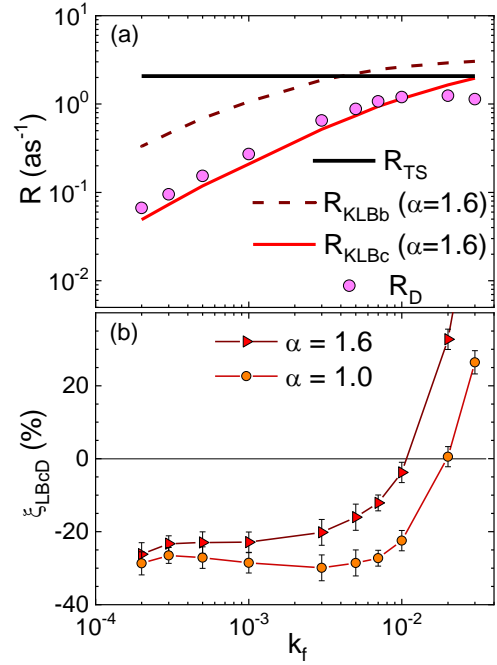

**Figure 3.** For the case of the deformation independent friction ($S_\eta = 0$), as functions of $k_f$ shown
(a) the fission rates: circles denote $R_D$; thick curve stands for the analytical rates $R_{KLB}$ calculated at $\alpha = 1.0$; the horizontal line indicates the friction independent transition state rate $R_{TS} = \omega(2\pi)^{-1} \exp(-U_b/\theta)$;
(b) the fractional differences $\xi_{LBD}$ corresponding to $\alpha = 1.0$ (circles), 1.3 (squares), and 1.6 (triangles).

**Figure 4.** For the case of the deformation-dependent friction ($S_\eta = 11$), as functions of $k_f$ shown
(a) the fission rates: circles denote $R_D$; thick dashed line stands for the analytical rates $R_{KLBb}$ calculated at $\eta = \eta_b$; thick solid curve denotes $R_{KLBc}$ calculated at $\eta = \eta_c$; the horizontal line indicates $R_{TS}$;
(b) the fractional difference $\xi_{LBcD}$ corresponding to $\alpha = 1.0$ (circles), and 1.6 (triangles).

In figure 3(b) we represent the fractional difference

$$\xi_{LBD} = \frac{R_{KLB}}{R_D} - 1. \tag{12}$$

Analyzing these values is more convenient because the rates themselves cover two orders of magnitude whereas the deviation of the approximate rate from the dynamical one is at most several tens of percent. One sees in figure 3(b) that the curves $\xi_{LBD}$ corresponding to different values of $\alpha$ converge as $k_f$ becomes smaller. This means that the analytical rate almost does not depend upon $\alpha$ here. For $k_f > 10^{-2}$, the fractional differences increase i.e. the agreement between $R_{KLB}$ and $R_D$ gets worse. This is not a surprise: here the applicability of the energy diffusion regime becomes poorer.

In average, $\xi_{LBD}$ with $\alpha = 1.0$ is the closest curve to zero in the appropriate range of $k_f$. That is why $R_{KLB}$ corresponding to this value of parameter $\alpha$ is shown in figure 3(a).

Due to the indications that the strength of friction is significantly reduced when the nuclear shape is compact, now we go over to the case of strong deformation dependence of friction (see figure 2). For this case, the time dependence of the dynamical fission rates is similar to those presented in figure 1 although the quasistationary values become different.

For the deformation dependent friction parameter, one can think of two modifications of equation (10). The first one results from using in equations (7), (8) the value of $\eta$ corresponding to the barrier, $\eta_b$. Let us denote this modified rate as $R_{KLBb}$; this is similar to what was done for the case of medium and strong friction (not considered in the present work). This rate is depicted in figure 4(a) by the dashed curve. One sees that this modification does not agree with $R_D$ at all: it overestimates $R_D$ by a factor of $2 \div 5$. Varying $\alpha$ does not remedy the situation.

The second possible modification is based on the fact that, in the energy-diffusion regime, before escaping over the barrier the particle crosses the well many times. Therefore, we try to use equations (7), (8) with the value of $\eta$ corresponding to the bottom of the well, $\eta_c$. The resulting rate, $R_{KLBc}$, is displayed in figure 4(a) by the thick solid curve. This $R_{KLBc}$ is in much better agreement with $R_D$. The quantitative comparison of the $R_{KLBc}$ with $R_D$ is shown in figure 4(b) using the fractional difference

$$\xi_{LBcD} = \frac{R_{KLBc}}{R_D} - 1. \tag{13}$$

In this case, there is more or less the same amount of agreement between $R_{KLBc}$ and $R_D$ for $\alpha = 1.0$ and 1.6. In general, the agreement of the approximate rate with the exact one is significantly worse ($\xi_{LBcD}$ is around $-25\%$) than for the case of the deformation independent friction ($|\xi_{LBD}|$ is smaller than 10%). We interpret this as evidence that the character of nuclear collective motion at $2 \cdot 10^{-4} < k_f < 10^{-2}$ is neither the pure phase space nor pure energy diffusion.

**4. Conclusions**
In the present work, we have compared an approximate analytical formula for the fission rate in the energy diffusion regime, i.e. at small values of friction, with the exact quasistationary numerical rate $R_D$. The latter was obtained modeling the fission process by means of the Langevin equations for the phase space. Our calculations demonstrate relatively good agreement between the approximate rate of equation (10) and $R_D$ provided the friction parameter $\eta$ is deformation independent.

For the case of the deformation dependent $\eta$, we have proposed a modification of approximate analytical quasistationary fission rate (see equations (7)-(11)). This modification consists of using the value $\eta(q_c)$. The difference between the modified approximate rate and $R_D$ does not exceed 30% for the given dependence $\eta(q)$. In our opinion, this comparatively large deviation occurs because in a single bounce of the fictious Brownian particle (fissioning trajectory) the energy diffusion regime becomes destroyed due to the sharp increase of $\eta$ with $q$.